\documentclass[structabstract]{aa}
\usepackage{graphicx}
\usepackage{txfonts}

\begin{document}
   \title{Bouncing Behavior of Microscopic Dust Aggregates}


   \author{A. Seizinger, \inst{1} \and W. Kley\inst{1}}

   \institute{Institut f\"ur Astronomie and Astrophysik, Eberhard Karls Universit\"at T\"ubingen,\\
              Auf der Morgenstelle 10c, D-72076 T\"ubingen, Germany\\
              \email{alexs@tat.physik.uni-tuebingen.de}\
             }

   \date{Received 18.12.2012; accepted 14.01.2013}

 
  \abstract
   {Bouncing collisions of dust aggregates within the protoplanetary may have a significant impact on the growth process of planetesimals. 
   Yet, the conditions that result in bouncing are not very well understood. 
   Existing simulations studying the bouncing behavior used aggregates with an artificial, very regular internal structure.}
   {Here, we study the bouncing behavior of sub-mm dust aggregates that are constructed applying different sample preparation methods.
    We analyze how the internal structure of the aggregate alters the collisional outcome and determine the influence
    of aggregate size, porosity, collision velocity, and impact parameter.}
   {We use molecular dynamics simulations where the individual aggregates are treated as spheres that are made up of several hundred thousand individual monomers.
     The simulations are run on GPUs.}
   {Statistical bulk properties and thus bouncing behavior of sub-mm dust aggregates depend heavily on the preparation method.
   In particular, there is no unique relation between the average volume filling factor and the coordination number of the aggregate.   
   Realistic aggregates bounce only if their volume filling factor exceeds $0.5$ and collision velocities are below $0.1\,\mathrm{m s^{-1}}$.}
   {For dust particles in the protoplanetary nebula we suggest that the bouncing barrier may not be such a strong handicap
    in the growth phase of dust agglomerates, at least in the size range of $\approx 100 \mu$m.
   }

   \keywords{Planets and satellites: formation -- Protoplanetary disks -- Methods: numerical}

   \authorrunning{Seizinger \& Kley}
   \maketitle
 
\section{Introduction}
For the planet formation process, the growth from micron sized dust grains to kilometer sized objects is a key ingredient of the core accretion scenario originally proposed by \citet{1996Icar..124...62P}. Yet, the question remains unanswered how this process is accomplished in the face of various impediments. First of all, fast inward drift limits the time available to form planetesimals by successive collisions to less than $10^4$ years \citep{1977MNRAS.180...57W}. The growth process itself heavily depends on two ingredients: 
\begin{enumerate}
 \item The dynamical properties of the disk that determine the collision rate as well as the parameters of a collision such as the impact velocity.
 \item The mechanical behavior of the colliding aggregates that determines the outcome of the collision.
\end{enumerate}
Since the information accessible through direct observations is limited the first aspect is addressed mainly by theoretical work and computer simulations \citep{2012MNRAS.420.2419F}.
For an overview on the properties of protoplanetary disks we refer to the following reviews by \citet{2007prpl.conf..555D} and \citet{2011ARA&A..49..195A}.

To investigate the collision behavior of dust/icy aggregates various methods are employed. Depending on the size of the aggregates and the desired collision velocity, laboratory experiments are possible. As of today, laboratory experiments provide data of collisions ranging from millimeter- to decimeter-sized aggregates composed of different materials (mainly Silicate/Quartz and Ice). A comprehensive summary of laboratory experiments is given by \citet{2008ARA&A..46...21B}. 
Computer simulations provide a second method to study the collisional behavior of dust or ice aggregates. 
Collisions of very small, micron sized aggregates have been simulated using a molecular dynamics approach \citep[e.g.\,][]{1997ApJ...480..647D, 2009A&A...507.1023P, 2007ApJ...661..320W, 2009ApJ...702.1490W}. For macroscopic aggregates different methods such as smoothed particle hydrodynamics (SPH) are employed \citep[e.g.\,][]{2007A&A...470..733S, 2010A&A...513A..58G}.

More recent experiments showed that collisions of mm to cm-sized aggregates often result in bouncing \citep[e.g.\,][]{2009ApJ...696.2036W, 2010Icar..206..424H, 2012Icar..218..688W, 2012A&A...542A..80J}. Extrapolating the results obtained from the various experiments \citet{2010A&A...513A..56G} devised a model describing the outcome of collision with respect to the collision velocity, and the mass and porosity of the colliding aggregates. Employing this model to simulate the evolution of a swarm of dust aggregates in a protoplanetary disk the so called ''bouncing barrier'' emerged \citep{2010A&A...513A..57Z}. As the aggregates grow larger their relative velocities increase. Due to the growing kinetic impact energy aggregates get increasingly compacted during successive collisions. When the aggregates get too compact their collisions do not result in sticking anymore. Instead, they bounce off each other and the growth process is stopped. This occurs in the size regime of centimeters.

A possible way to overcome the bouncing barrier has been recently suggested by \citet{2012A&A...540A..73W}. Under the assumption of a few bigger aggregates that act as initial seeds it is possible to grow larger $100\,\mathrm{m}$ sized bodies on the timescale of $1\,\mathrm{Myr}$. A possible origin of those seeds has been proposed by \citet{2012A&A...544L..16W}. Taking into account a Maxwellian velocity distribution they found that low velocity collisions can allow a few aggregates to grow considerably larger than the average of the simulated population.

Despite its significant influence on the growth process, bouncing still lacks theoretical understanding of its prerequisites on a microscopical scale. So far, molecular dynamics (MD) simulations result in bouncing only for rather compact aggregates \citep{2011ApJ...737...36W, 2012ApJ...758...35S}. According to \citet{2011ApJ...737...36W} an average coordination number of $6$ is required for aggregates to bounce off each other. However, in laboratory experiments bouncing frequently observed in collisions of aggregates with lower filling factors / coordination numbers for which MD simulations clearly predict sticking. It has been speculated that this discrepancy could result from a size effect or a possible compaction of the outer shell during the handling process of the aggregates used in the laboratory experiments. However, the latter hypothesis has been refuted by recent experiments (Kothe et al. 2012).

The aim of this work is to study the influence of the internal structure on the bouncing behavior of sub-mm dust aggregates in greater detail. Using the enormous computing power provided by GPUs we can simulate aggregates consisting of several hundreds of thousands of monomers and thus simulate aggregates in a size range from several microns up to $\approx 0.1\,\mathrm{mm}$ in diameter.

\section{Interaction model}
\label{sec:interaction_model}

To simulate the behavior of of dust aggregates we use a soft sphere discrete element method (SSDEM) approach. The dust aggregates are composed of hundreds of thousands of micron sized spherical grains (monomers). Our interaction model is based on the work of \citet{1997ApJ...480..647D} who combined earlier theoretical work by \citet{1971RSPSA.324..301J,1995PMagA..72..783D, 1996PMagA..73.1279D} into a detailed micro-mechanical model describing the interaction between two monomers. These monomers may establish adhesive contacts when touching each other and kinetic energy is dissipated upon deformation of these contacts. A few years later, \citet{2007ApJ...661..320W} presented a different approach when deriving nearly the same forces and torques from corresponding potentials. 

When trying to reproduce the results of laboratory experiments performed by \citet{2009ApJ...701..130G} on the compression of porous dust cakes \citet{2012A&A...541A..59S} observed that the behavior predicted by the model of \citet{1997ApJ...480..647D} was too soft. Since the samples used by \citet{2009ApJ...701..130G} had been composed of micron-sized, spherical, monodisperse  silicate grains their results constituted a perfect possibility to calibrate the model. Introducing two free parameters $m_{\mathrm{r}}$ and $m_{\mathrm{s}}$ that modify the strength of the rolling and sliding interaction between two monomers \citet{2012A&A...541A..59S} were able to obtain excellent agreement between laboratory results and computer simulations.

In this work we use the modified interaction model presented by \citet{2012A&A...541A..59S} with $m_{\mathrm{r}} = 8$ and $m_{\mathrm{s}} = 2.5$. The material parameters are listed in Tab.\,\ref{tab:material_parameters}.

\begin{table}
 \caption[]{Material Parameters of the individual monomers used in the simulations.}
 \label{tab:material_parameters} 
 \centering
 \renewcommand\arraystretch{1.2}
 \begin{tabular}{ll}
   \hline
   \noalign{\smallskip}
   Physical property & Silicate\\
   \noalign{\smallskip}
   \hline
   \noalign{\smallskip}
   Particle Radius $r$ (in $\mathrm{\mu m}$) & $0.6$\\
   Density $\rho$ (in g\, cm$^{-3}$) & $2.65$\\
   Surface Energy $\gamma$ (in mJ\, m$^{-2}$) & $20$\\
   Young's Modulus $E$ (in GPa)         & $54$\\
   Poisson Number $\nu$               & $0.17$\\
   Critical Rolling Length $\xi_\mathrm{crit}$ (in nm) & $2$\\
   \noalign{\smallskip}
   \hline
 \end{tabular}
\end{table}

\section{Sample generation}
\label{sec:sample_generation}
\begin{figure*}
\resizebox{\hsize}{!}{\includegraphics{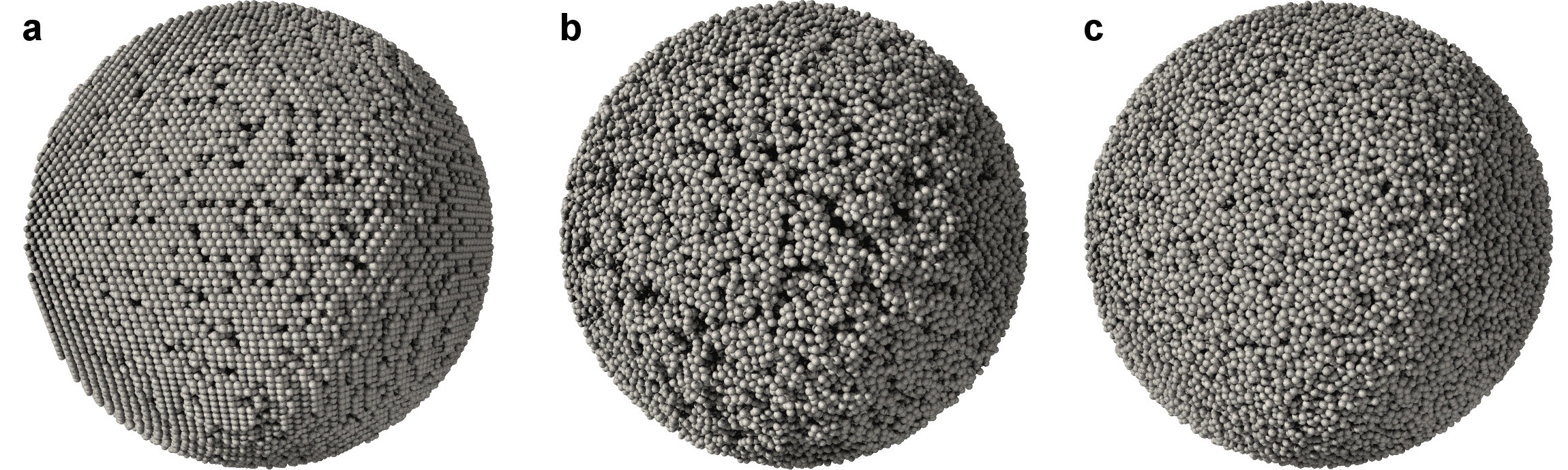}}
\caption{Examples of the different types of aggregates used in this work: (a) Hexagonal lattice $\phi = 0.59, n_\mathrm{c} = 9.93$, (b) Ballistic aggregation with migration $\phi = 0.40, n_\mathrm{c} = 3.98$, and (c) Static compaction $\phi = 0.49, n_\mathrm{c} = 3.50$. All depicted spheres have a diameter of $60\,\mathrm{\mu m}$.} 
\label{fig:aggregates}
\end{figure*}

In this work we examine the conditions under which bouncing occurs. Apart from the external parameters describing the physics of collisions such as the impact parameter or velocity we study the influence of the internal structure of the aggregate. Examples of such aggregates that have been generated by different methods are shown in Fig.\,\ref{fig:aggregates}. 

To study the influence of the aggregate size we use aggregates with diameters in the range of $30$ to $100\,\mathrm{\mu m}$. Unfortunately, simulations with larger aggregates are infeasible due to the required computational cost, at least for a wider range of parameters.

Typically, the volume filling factor $\phi$ and the average coordination number $n_\mathrm{c}$ are used to classify aggregates. In general, the filling factor is given by \begin{eqnarray}\phi = \frac{N V_\mathrm{p}}{V_\mathrm{A}}  ,\end{eqnarray}
where $N$ denotes the number of monomers, $V_\mathrm{p}$ is the volume of a monomer, and $V_\mathrm{A}$ is the total volume occupied by the aggregate. As we use spherical aggregates $V_\mathrm{A}$ can be calculated easily from the outer radius of the aggregate. 
For irregular shaped aggregates there are different possibilities to define $V_\mathrm{A}$. For example \,\citet{1992A&A...263..423K} determine the size of a porous aggregate based on its radius of gyration whereas \citet{1993A&A...280..617O} use the geometric cross section. It is important to keep this ambiguity in mind when comparing the porosities of fluffy aggregates to other works.

In molecular dynamics simulations, the coordination number of a monomer denotes the number of the neighbors it interacts with. Thus, the average coordination number $n_\mathrm{c}$ is obtained by averaging the number of contacts of each particle.

In this work we use three different types of aggregates: Hexagonal lattice aggregates featuring a regular grid structure, aggregates produced by slowly compacting a porous dust cake, and aggregates generated by successively adding single monomers from randomly chosen directions. These choices have been motivated by the fact that hexagonal lattice aggregates are easy to build and allow for comparison with earlier work by \citet{2011ApJ...737...36W}, whereas the static compaction resembles the generation of samples used in laboratory results. The aggregates of the third type are generated algorithmically but their structure remains comparable to the static compaction type (see Sect.\,\ref{sec:comparison}). 

Because the aggregates within the protoplanetary nebula grow through successive collisions one might expect that their internal structure lies somewhere in between the static compression and the ballistic aggregation cases.

\subsection{Hexagonal lattice (CPE)}
Hexagonal-lattice type aggregates (also referred to as hexagonal close packing with extraction (CPE)) may be generated very easily. First, a hexagonal close packing aggregate is generated which features a volume filling factor $\phi \approx 0.74$ and a coordination number $n_\mathrm{c} \approx 12$ (due to surface effects $n_c$ equals $12$ only for aggregates of infinite size). In the second step a suitable number of randomly selected monomers will be removed to achieve the desired volume filling factor. As a result a small number of monomers on the surface may be become disconnected from the main aggregate and will be removed as well.

\citet{2011ApJ...737...36W} have already studied the bouncing behavior of this type of aggregates and found that bouncing will occur if the average coordination number is greater than $6$. \citet{2012ApJ...758...35S} analyzed the relation between the coefficient of restitution and the collision velocity in experiments and simulations using CPE-aggregates. Their results agreed well with a theoretical model by \citet{Thornton1998154}.

\subsection{Ballistic aggregation with migration (BAM)}
\label{sec:bam}
\begin{figure}
\resizebox{\hsize}{!}{\includegraphics{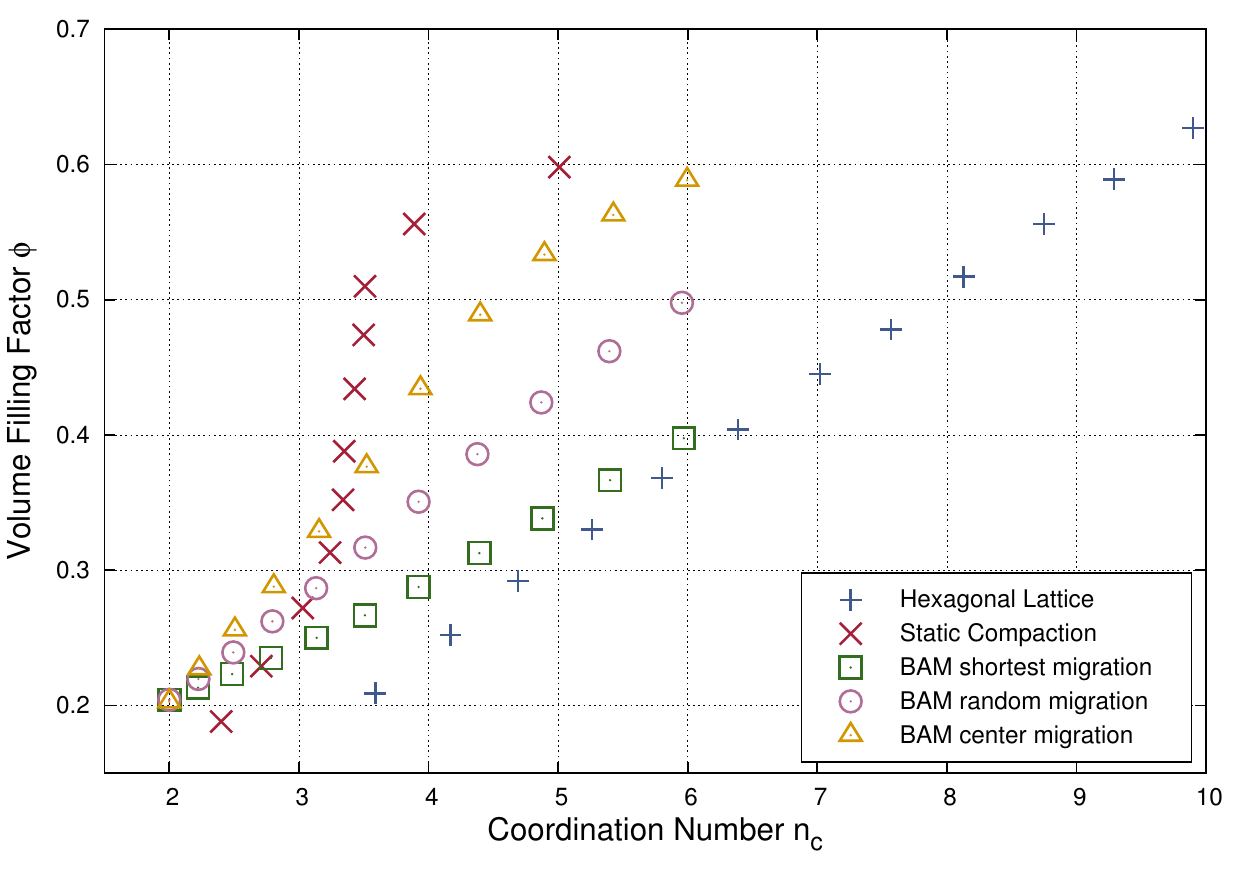}}
\caption{The relation between the volume filling factor $\phi$ and the average coordination number $n_\mathrm{c}$. All aggregates are spherical and have a diameter of $60\,\mathrm{\mu m}$.}
\label{fig:ff_vs_nc}
\end{figure}

The second type of aggregates was originally suggested by \citet{2008ApJ...689..260S} and also studied in the work of \citet{2011ApJ...737...36W}. To generate a larger aggregate single monomers are successively shot in from random directions onto the existing aggregate. When the monomer hits the aggregate it will either remain at the position where the first contact has been established or migrate to a position close by where it establishes contacts with two or three monomers. Compared to \citet{2008ApJ...689..260S}, we use three different methods to select the final position of the migrating particle:
\begin{enumerate}
 \item Select the position closest to the spot, where the monomer impacts on the aggregate (referred to as ``shortest migration'').
 \item Select the position randomly from all available possibilities (referred to as ``random migration'').
 \item Select the position which is closest to the center of mass (referred to as ``center migration'').
\end{enumerate}
For a given coordination number the resulting aggregates show a different filling factor depending on which selection mechanism is employed (see Fig.\,\ref{fig:ff_vs_nc}). The first method leads to rather porous aggregates since the monomers typically migrate to positions further outward compared to the case of random migration. Likewise, the resulting aggregates will become even more compact if monomers migrate to the most inward position available.

Note that we do not claim that random or center migration are realistic growth processes that accurately describe the growth of dust aggregates in protoplanetary discs. Yet, they constitute a computationally very cheap approach to generate larger aggregates that do not suffer from the artificial lattice structure like the CPE aggregates described above. Compared to the ``static compaction''-aggregates they additionally offer the advantage that they are perfectly relaxated. Since all monomers are in equilibrium distance from each other, there are no attractive or repulsive forces that could lead to a breakup of the aggregate.

\subsection{Static compaction (SC)}
The last type of aggregates used for our the studies is the most computationally expensive. To generate a spherical aggregate of a certain diameter we start with a sufficiently large, cuboid shaped dust cake generated by random ballistic deposition (RBD). Since RBD-aggregates feature an initial volume filling factor of $\phi = 0.15$ we first have to compact the aggregate until we reach the desired filling factor. For this purpose, the aggregate is put into a box of walls that may move towards each other. According to \citet{2012A&A...541A..59S} this compaction must be very slow to avoid inhomogeneities. 

Even if the cake is compacted homogeneously for filling factors above $\approx 0.45$ it will get elastically charged and expand when the compacting walls are removed. Therefore the dust cakes needs to be  relaxated before removing the walls of the compaction box. For this purpose we disturb the aggregate by modifying the position of the monomers randomly by a very small amount. We keep the disturbed aggregate in a box of fixed size and wait until the kinetic energy induced by the disturbance is damped away by the inelastic monomer interaction. To get rid of kinetic energy below the threshold where the inelastic regime is entered we additionally enforce a viscous damping mechanism. For this purpose, the velocities and angular velocities of the monomers are multiplied by a factor of $1-\kappa$ in each time step, where $\kappa$ denotes a damping coefficient. In this work we use $\kappa = 0.0001$.

It turns out that a maximum disturbance of a factor of $0.001$ of the radius of a monomer is sufficient to stabilize the aggregate without altering its internal structure. Higher values may alter the coordination number significantly which could potentially change the collisional behavior of the aggregates and is therefore unwanted. For fillings factors above $\phi = 0.58-0.59$ this method does not work anymore. Here the compaction is too close to the random closest packing ($\phi \approx 0.63$) and no stable configuration can be reached without rearranging the monomers significantly.

After the aggregate has been relaxated the compaction box is removed and a spherical aggregate will be cut out of the compacted cake. As this procedure is computationally very expensive it takes several days to generate larger ($50\, \mathrm{\mu m}$ in diameter and above) aggregates of this type. 

\subsection{Comparison}
\label{sec:comparison}
Comparing the relation between $\phi$ and $n_\mathrm{c}$ of the different types of aggregates described above shows very interesting features: As we can see in Fig.\,\ref{fig:ff_vs_nc} the relation depends considerably on the preparation method. The different BAM generation methods have been described in Sect.\,\ref{sec:bam}. 

From the work presented in this section two important conclusions can be drawn:
\begin{enumerate}
 \item The coordination number $n_\mathrm{c}$ is not sufficient to describe the properties of an aggregate. Especially, there is no unambiguous relation $\phi(n_\mathrm{c})$ between filling factor and coordination number. 
 \item Hexagonal lattice (CPE) aggregates have a very distinct relation $\phi(n_\mathrm{c})$ compared to the other two methods that produce aggregates with less artificial structures. 
\end{enumerate} 
In laboratory experiments with aggregates composed of micron sized dust grains, it is typically only possible to determine the filling factor but not the coordination number. Thus, one has to be very careful when comparing results from numerical simulations of CPE aggregates with laboratory experiments.

\section{Results}
\label{sec:results}
In this section we present our results from various simulations in which we study the influence of the collisions velocity, impact parameter, and aggregate size on the bouncing behavior. All simulations have been performed on NVIDIA GPUs (GTX460, GTX570, Tesla C2070). Depending on the aggregate size and filling factor each simulation took between less than an hour and half a day. 

\subsection{Growth Factor}
In the following bouncing maps the ``growth factor'' $\gamma$ that is inspired by the four-population model suggested by \citet{2011A&A...531A.166G} is depicted. It is defined by 
\begin{eqnarray}\gamma = \frac{m_{\mathrm{largest}}}{m_{\mathrm{tot}}}, \label{eq:gamma} \end{eqnarray}
where $m_{\mathrm{largest}}$ is the mass of the largest fragment and $m_{\mathrm{tot}}$ the total mass of the colliding aggregates. For perfect sticking we obtain $\gamma = 1$, for total destruction $\gamma \rightarrow 0$. In collisions of equal sized aggregates, a value $\gamma = 0.5$ indicates bouncing. However, during the transition from perfect sticking to fragmentation $\gamma$ may also become $0.5$. To distinguish between the two cases we consider the mass ratio $\gamma_2$ of the second largest fragment. In the bouncing case it is $0.5$ as well whereas in the fragmentation case the mass of the second largest fragment is much lower than 0.5 of the total mass as there are a lot of other smaller fragments.

Thus, in the bouncing maps presented in this work the green areas indicate sticking, the upper left yellow areas bouncing, and the color gradient from green to yellow to red on the right marks the transition from sticking to fragmentation.

Note that $\gamma = \gamma_2 = 0.5$ only applies in the case of ``perfect bouncing''. In our simulations we often observe the loss of a few monomers (typically less than $100$) which is negligible compared to the total number of monomers of $5 \cdot 10^4$ to $5 \cdot 10^5$. Thus, we also count collisions as bouncing events if $\gamma$ and $\gamma_2$ are slightly smaller than $0.5$.

\subsection{Hexagonal lattice}

\begin{figure}
\resizebox{\hsize}{!}{\includegraphics{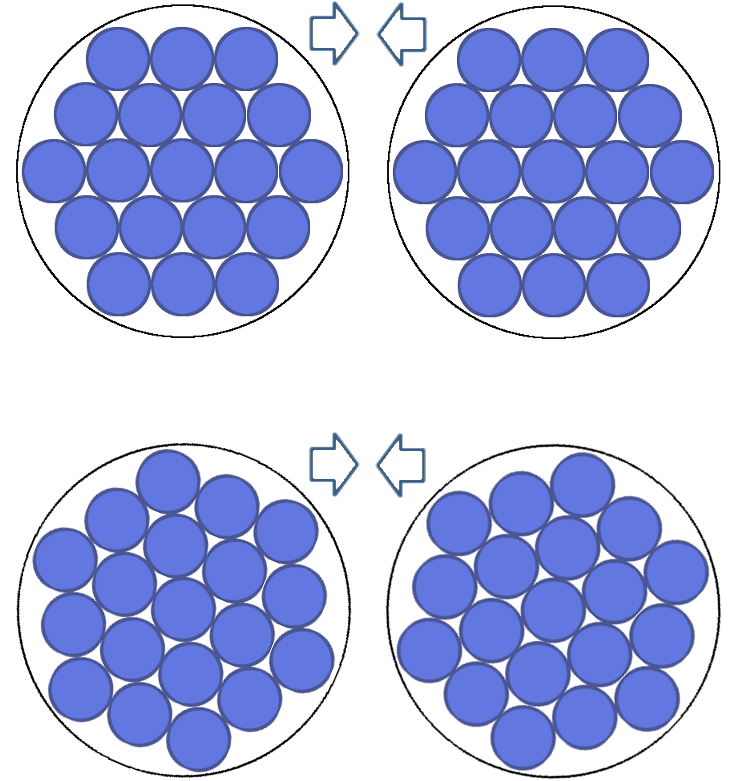}}
\caption{ Schematic view of the possible alignment of two CPE aggregates. In the upper case the aggregates are oriented with respect to their lattice structure. In the lower case the orientation is arbitrary.}
\label{fig:alignment}
\end{figure}

\begin{figure*}
\resizebox{0.50\hsize}{!}{\includegraphics{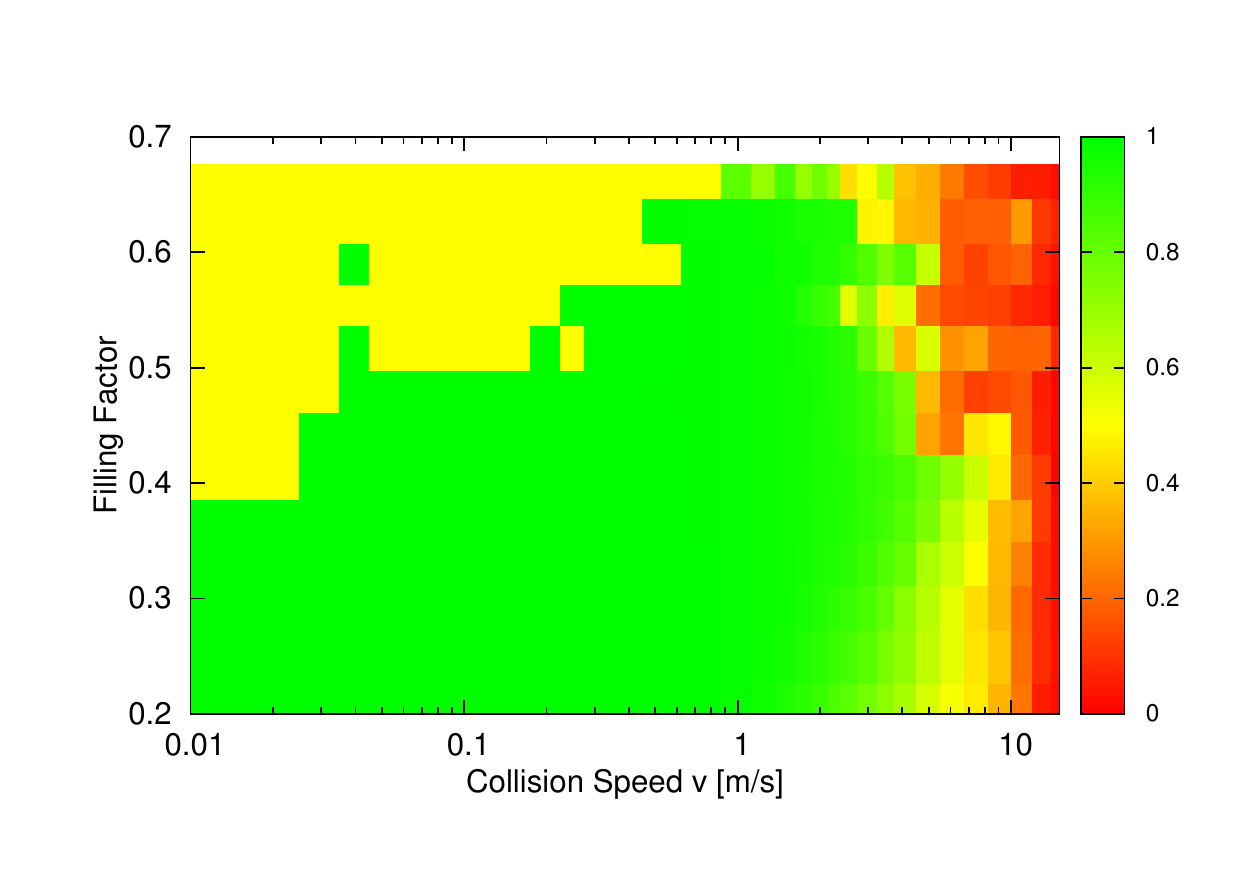}} \hfill \resizebox{0.50\hsize}{!}{\includegraphics{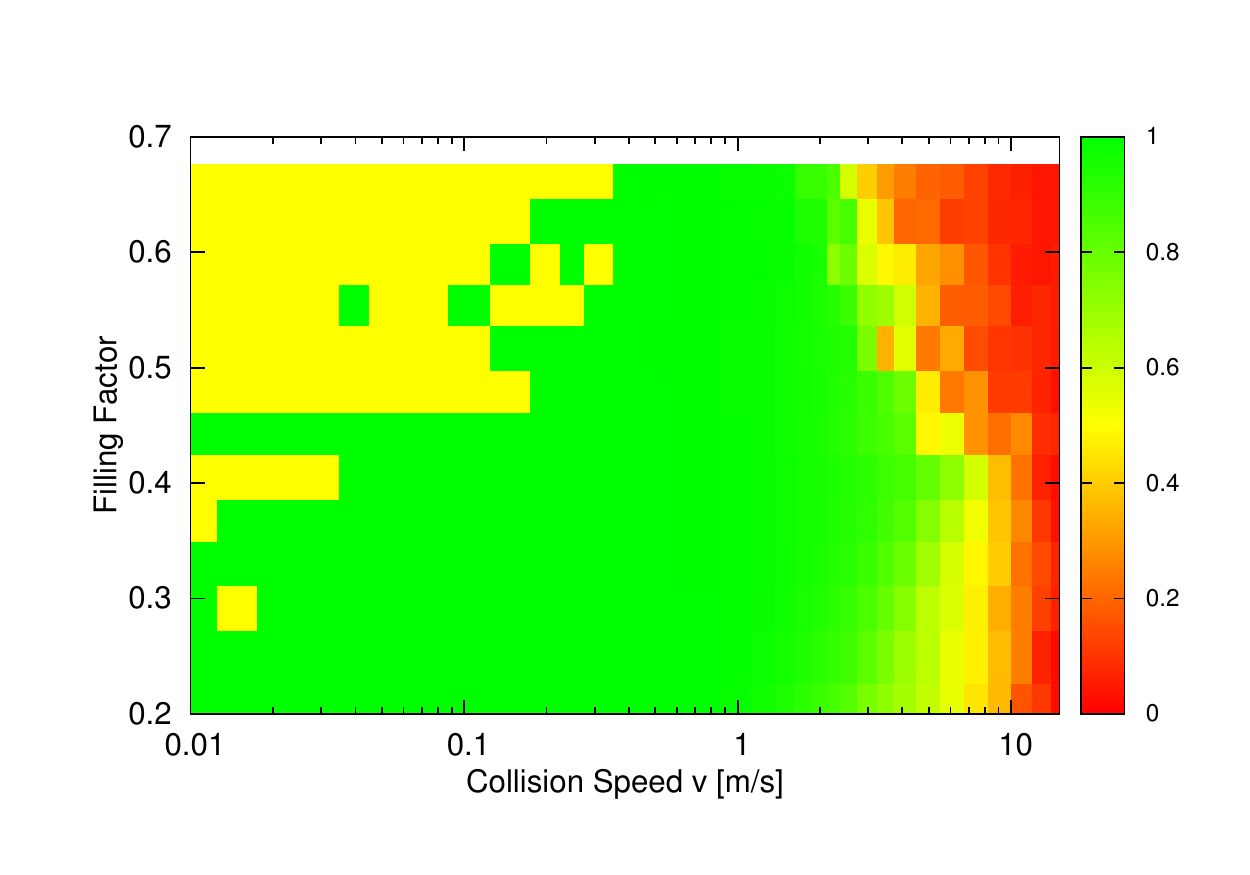}}
\caption{Growth factor, $\gamma$ (eq.~\ref{eq:gamma}), of the collision of two CPE aggregates with a diameter of $60\,\mathrm{\mu m}$. Sticking occurs in the green colored area whereas the yellow area in the upper left indicates bouncing. \textbf{Left:} Orientation aligned to the lattice structure of the aggregates. \textbf{Right:} Non aligned orientation.}
\label{fig:hex_orientation_dependency}
\end{figure*}

The outcome of head-on collisions of CPE aggregates has already been studied by \citet{2011ApJ...737...36W} who observed bouncing if the coordination number was greater than $6$. However, their aggregates were much smaller ($\approx 10^4$ monomers). As hexagonal lattice aggregates feature a regular lattice structure their orientation is likely to influence the collision behavior. Thus, we first examine the effect the orientation by comparing the case where the aggregates are aligned to their lattice structure (see upper part of Fig.\,\ref{fig:alignment}) to a random orientation (lower part of Fig.\,\ref{fig:alignment})). As we can see in Fig.\,\ref{fig:hex_orientation_dependency}, the orientation of the aggregates is important especially for the transition from sticking to bouncing with increasing filling factor. Looking at the left panel of Fig.\,\ref{fig:hex_orientation_dependency} and comparing the filling factor with the coordination number in Fig.\,\ref{fig:ff_vs_nc} we can reproduce the $n_\mathrm{c} \geq 6$-criterion proposed by \citet{2011ApJ...737...36W} for the aligned case. On the other hand, the bouncing maps differs significantly for a non aligned orientation (see right panel of Fig.\,\ref{fig:hex_orientation_dependency}).

\begin{figure*}
\resizebox{0.50\hsize}{!}{\includegraphics{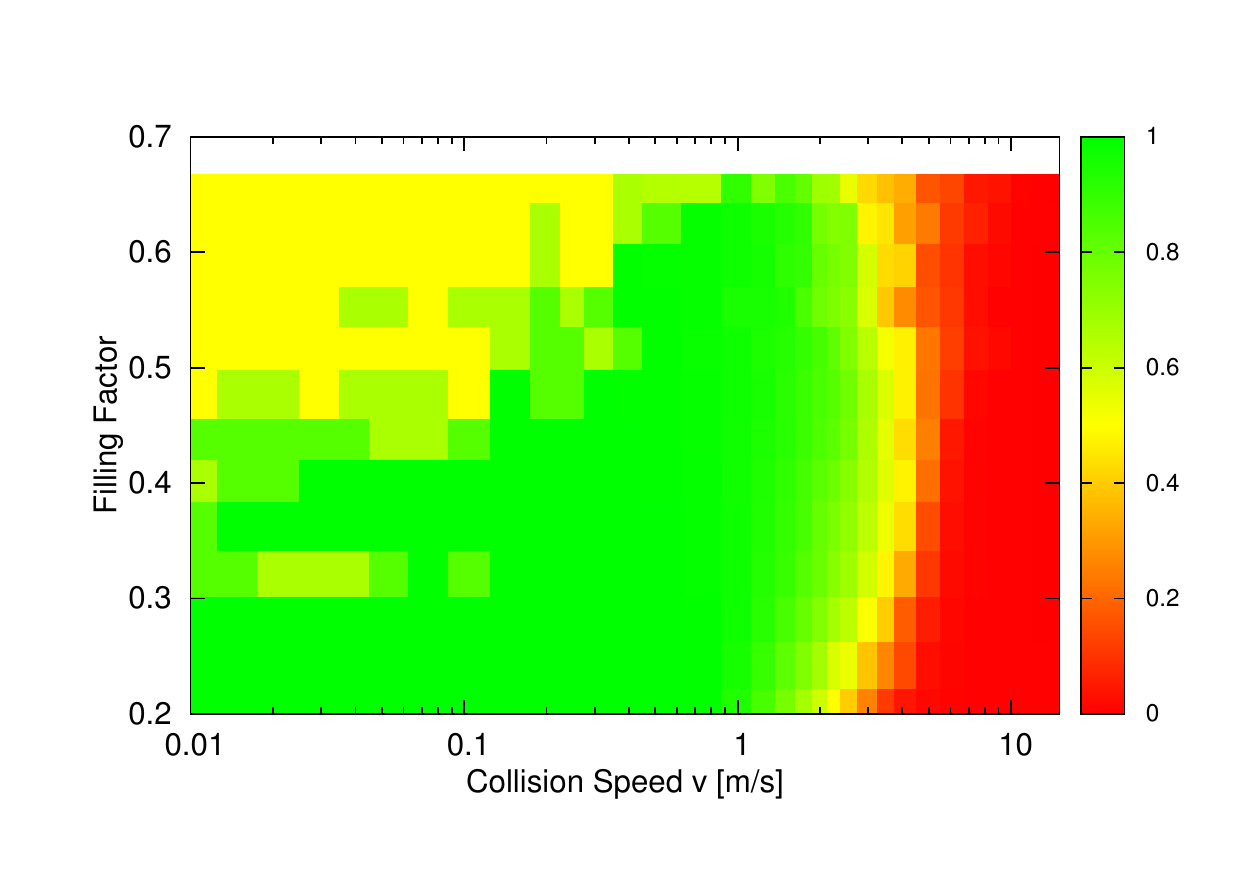}} \hfill \resizebox{0.50\hsize}{!}{\includegraphics{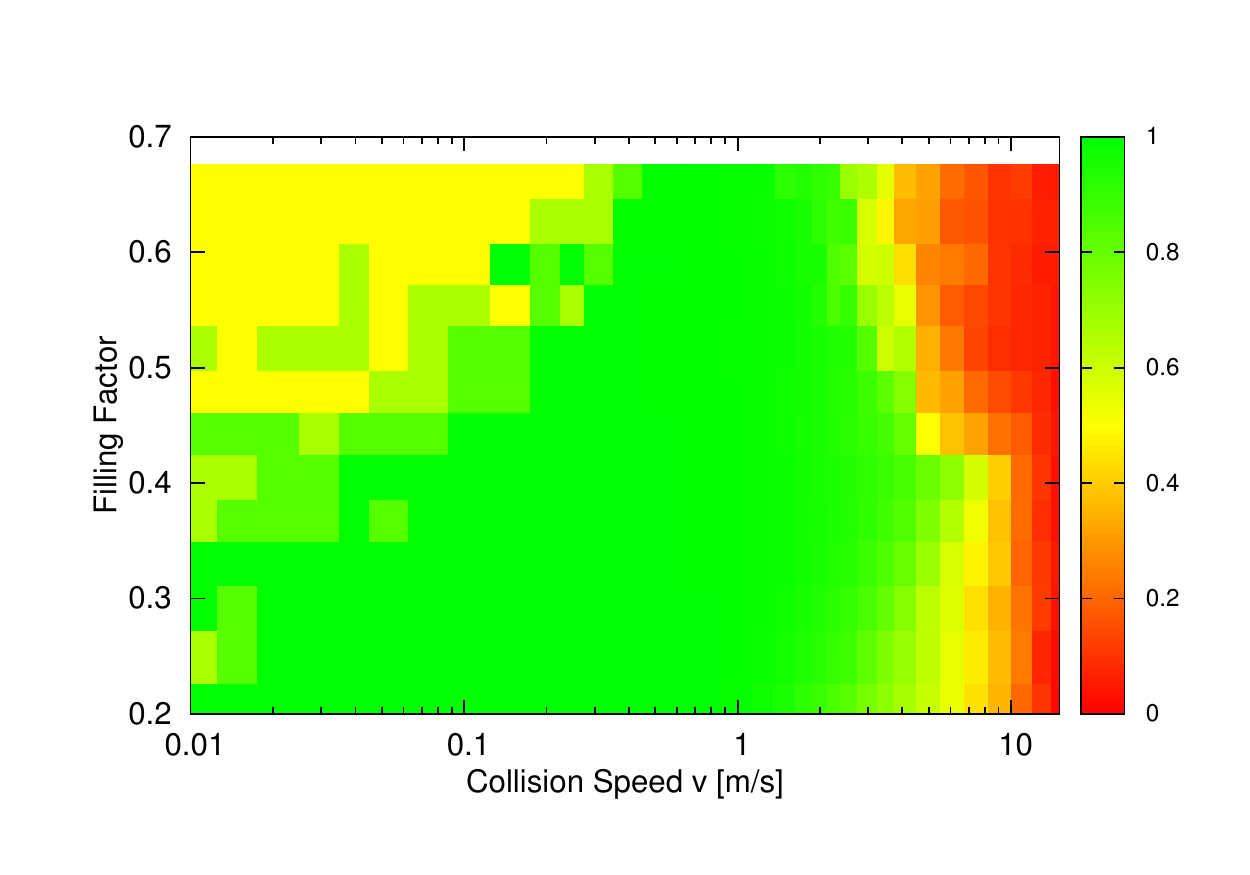}}
\caption{ Growth factor of the collision of two CPE aggregates of different size averaged over three different orientations. \textbf{Left:} Aggregates of a diameter of $30\,\mathrm{\mu m}$. \textbf{Right:} Aggregates of a diameter of $60\,\mathrm{\mu m}$.}
\label{fig:hex_size_dependency}
\end{figure*}

In order to mitigate the effect of the orientation we averaged over three different orientations to investigate the size dependency of our results. Each map has been generated using $12$ different filling factors and $28$ velocities. Thus, $3 \cdot 336 = 1008$ simulations had to be performed in total. Concerning bouncing we could not observe a clear difference between aggregates with a diameter of $30$ and $60\,\mathrm{\mu m}$ (see Fig.\,\ref{fig:hex_size_dependency}).

However, the velocity $v_{\mathrm{s}\rightarrow\mathrm{f}}$ at which the transition from sticking to fragmentation occurs changes significantly. For the small aggregates we get $v_{\mathrm{s}\rightarrow\mathrm{f}} \approx 4\,\mathrm{m s^{-1}}$ (left panel of Fig.\,\ref{fig:hex_size_dependency}). For the bigger aggregates we observe that $v_{\mathrm{s}\rightarrow\mathrm{f}}$ depends on the filling factor. For $\phi < 0.43$ we get $v_{\mathrm{s}\rightarrow\mathrm{f}} \approx 10\,\mathrm{m s^{-1}}$ whereas $v_{\mathrm{s}\rightarrow\mathrm{f}} \approx 5\,\mathrm{m s^{-1}}$ for $\phi > 0.43$ (right panel of Fig.\,\ref{fig:hex_size_dependency}). This can be explained by the reduced capability of compact aggregates to dissipate kinetic energy by restructuring. Taking into account Fig.\,\ref{fig:ff_vs_nc} we see that the transition occurs when the average coordination number $n_\mathrm{c}$ exceeds a value of $6$. A monomer with six or more contacts is fixated rather tightly and and thus the aggregate cannot change its internal structure as easily anymore.

In summary it can be said that for hexagonal lattice aggregates we regularly observe bouncing collisions for filling factors above $0.5$ and collision velocities up to roughly $0.2 \mathrm{m s^{-1}}$.

\subsection{Ballistic aggregation with migration}
\label{sec:results_bam}
\begin{figure}
\resizebox{\hsize}{!}{\includegraphics{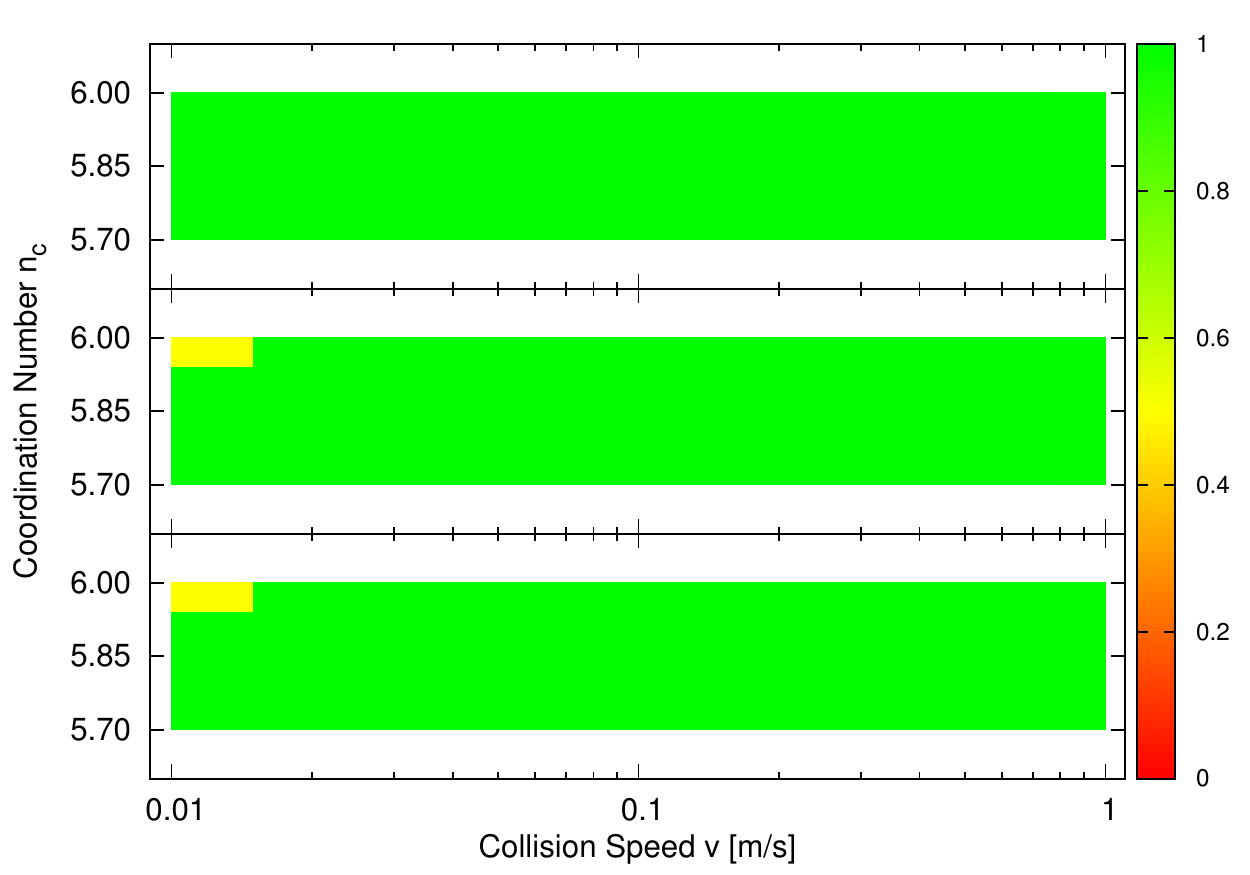}}
\caption{ Growth factor of the collision of two BAM aggregates generated by the shortest migration method. For three different aggregates of a diameter of roughly $75\,\mathrm{\mu m}$ only two collisions at $1\,\mathrm{cm s^{-1}}$ lead to bouncing.}
\label{fig:bam_shortest}
\end{figure}

\citet{2011ApJ...737...36W} found that bouncing may occur if $n_\mathrm{c} \geq 6$ independent of the type of aggregate they used. In Fig.\,\ref{fig:bam_shortest}, we show the outcome of collisions between two roughly $75\,\mathrm{\mu m}$ sized BAM aggregates generated by using the shortest migration method described in Sect.\,\ref{sec:bam}. The corresponding filling factor is between $0.36$ and $0.39$.  However, we did observe only two bouncing collisions. Since $n_\mathrm{c} = 6$ is the maximum value that can be achieved by two times migration we could not investigate what happens at higher coordination numbers.

Repeating the setup described above for the center migration case we get similar results as for the shortest migration case shown in Fig.\,\ref{fig:bam_shortest}. We observe hardly any bouncing events even for aggregates with $n_\mathrm{c} = 6$ (which corresponds to a filling factor of $0.49\,-\,0.5$). This indicates that the bouncing behavior of BAM aggregates depends more on the filling factor than the coordination number.

\begin{figure*}
\resizebox{0.50\hsize}{!}{\includegraphics{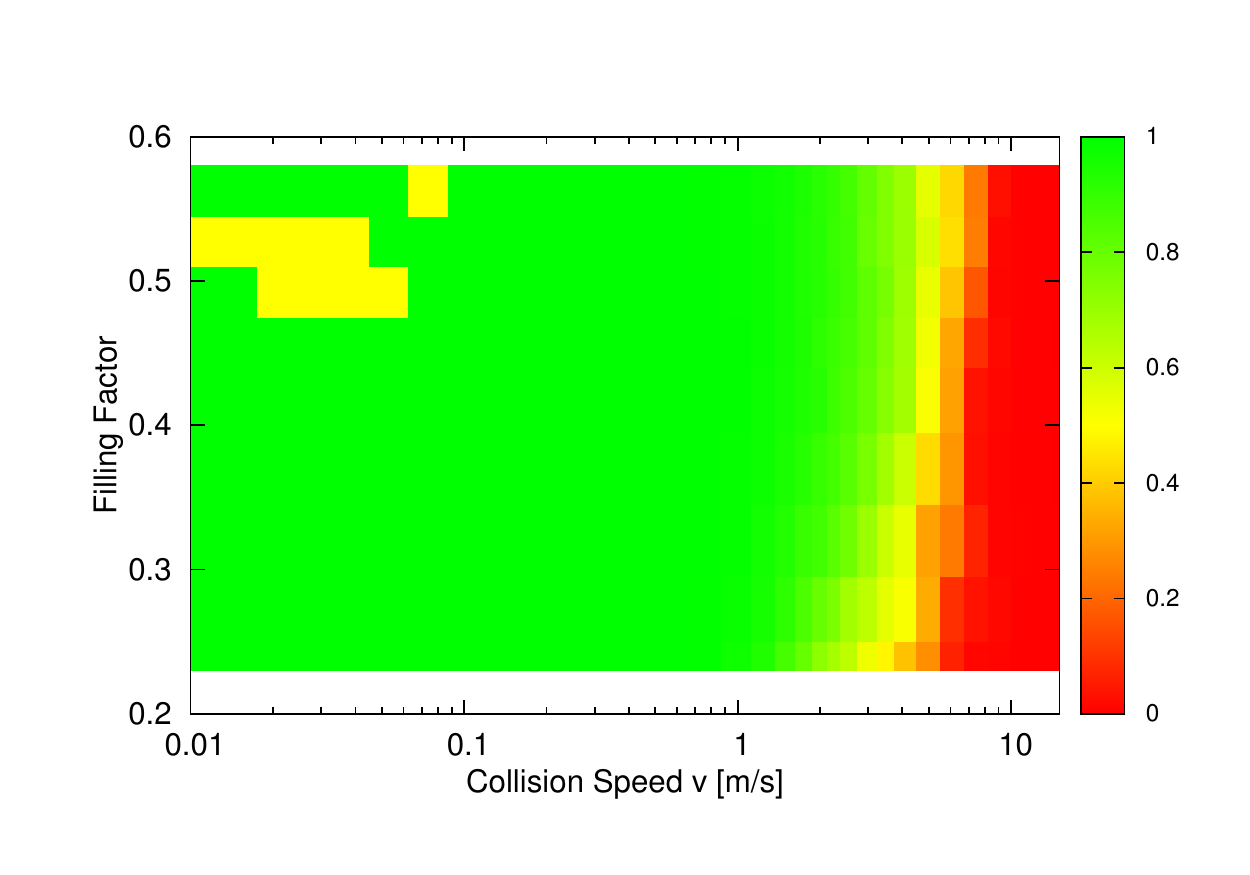}} \hfill \resizebox{0.50\hsize}{!}{\includegraphics{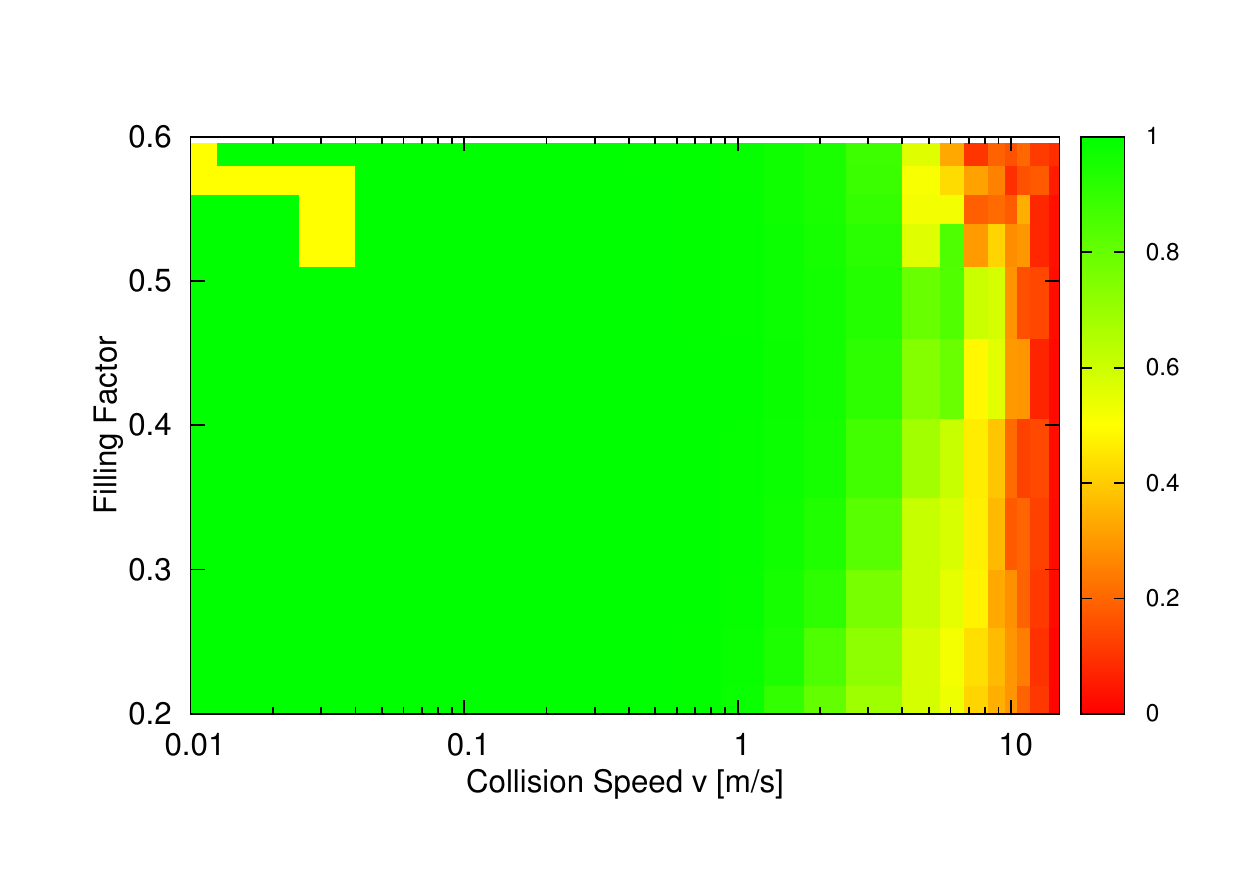}}
\caption{Growth factor of the collision of two differently sized BAM aggregates that have been generated with the center migration method. \textbf{Left:} Diameter of $30\,\mathrm{\mu m}$. \textbf{Right:} Diameter of $60\,\mathrm{\mu m}$.}
\label{fig:bam_size_dependency}
\end{figure*}

To achieve higher filling factors we switched to the center migration method (see Sect.\,\ref{sec:bam}). The corresponding bouncing maps are shown in Fig.\,\ref{fig:bam_size_dependency}. It is striking that the bouncing regime is much smaller compared to the CPE aggregates. As before, for larger aggregates the transition from sticking to fragmentation occurs at higher velocities.

\subsection{Static compaction}
\begin{figure*}
\resizebox{0.50\hsize}{!}{\includegraphics{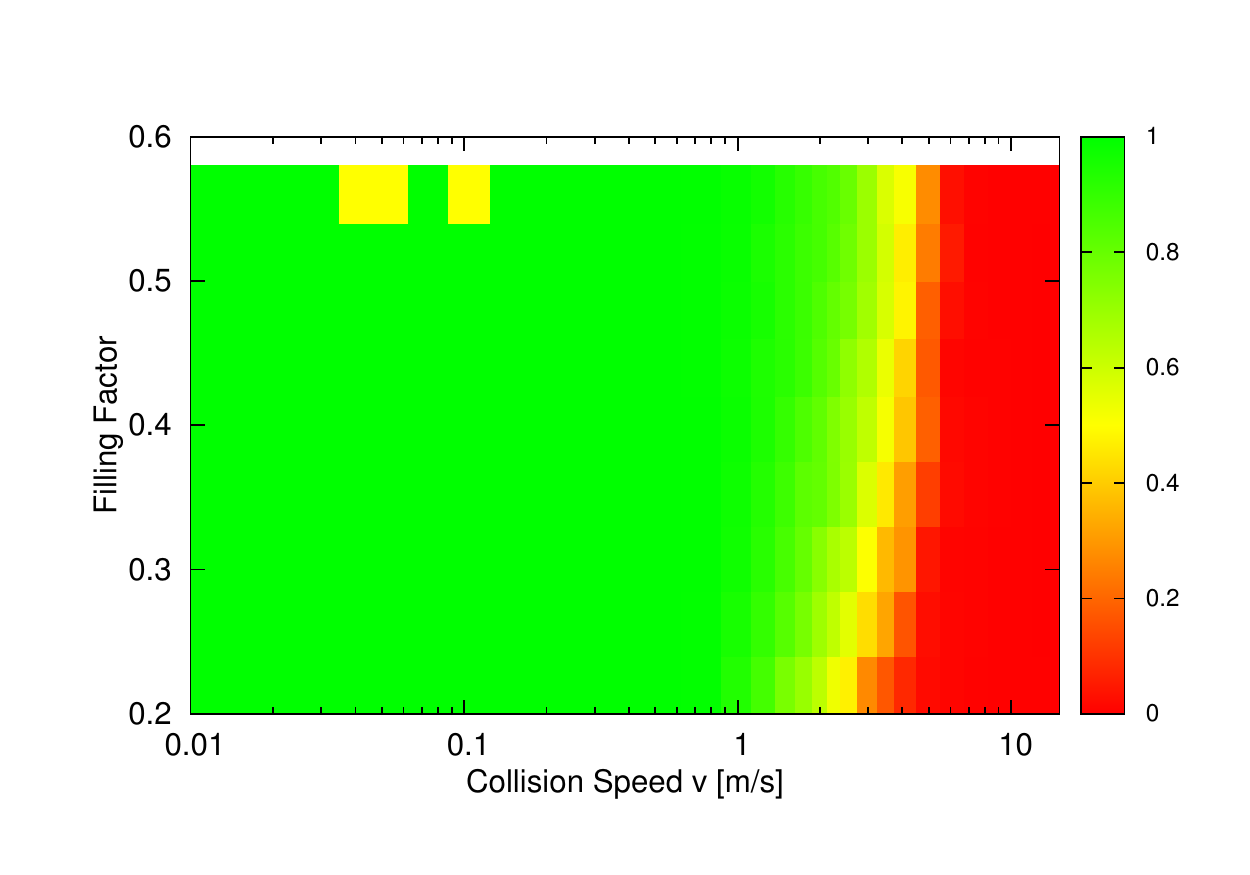}} \hfill \resizebox{0.50\hsize}{!}{\includegraphics{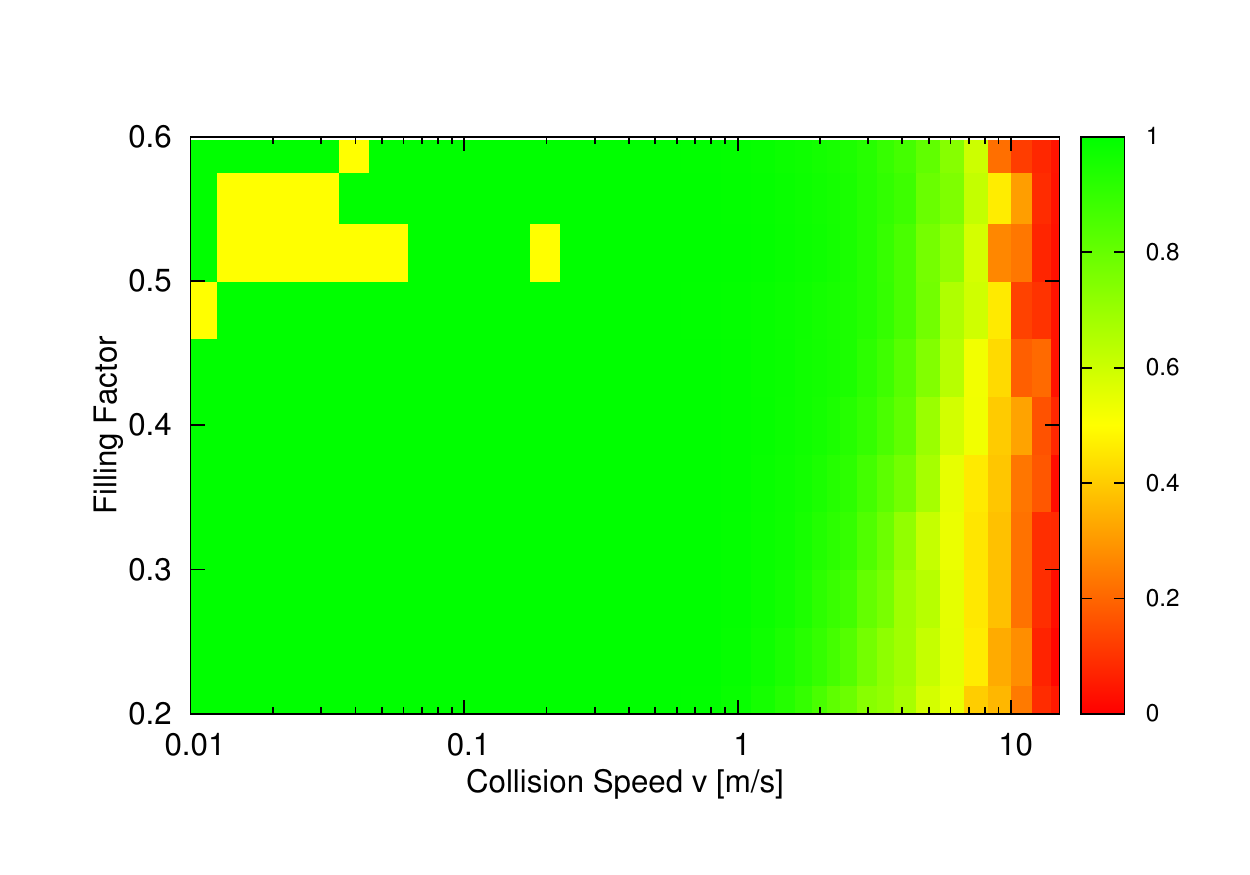}}
\caption{Growth factor of the collision of two static compaction aggregates of different size. \textbf{Left:} Diameter of $30\,\mathrm{\mu m}$. \textbf{Right:} Diameter of $60\,\mathrm{\mu m}$.}
\label{fig:irr_size_dependency}
\end{figure*}

The bouncing behavior of the aggregates generated by static compaction is quite similar to the BAM aggregates (see Fig.\,\ref{fig:irr_size_dependency}). Again, the bouncing regime is considerably smaller compared to the case of hexagonal lattice aggregates and bouncing is observed only in some cases for high filling factors above $0.5$ and collision velocities below $0.1\,\mathrm{m s^{-1}}$.

As in the case of the other aggregate types the transition velocity $v_{\mathrm{s}\rightarrow\mathrm{f}}$ from sticking to fragmentation increases with increasing aggregate size. For the the small aggregates with $d = 30\,\mathrm{\mu m}$ we observe $v_{\mathrm{s}\rightarrow\mathrm{f}} \approx 4\,\mathrm{m s^{-1}}$ whereas for $d = 60\,\mathrm{\mu m}$ the transition velocity goes up to $v_{\mathrm{s}\rightarrow\mathrm{f}} \approx 12\,\mathrm{m s^{-1}}$.

\subsection{Size dependency}
\label{sec:size_dependency}

\begin{figure*}
\resizebox{0.50\hsize}{!}{\includegraphics{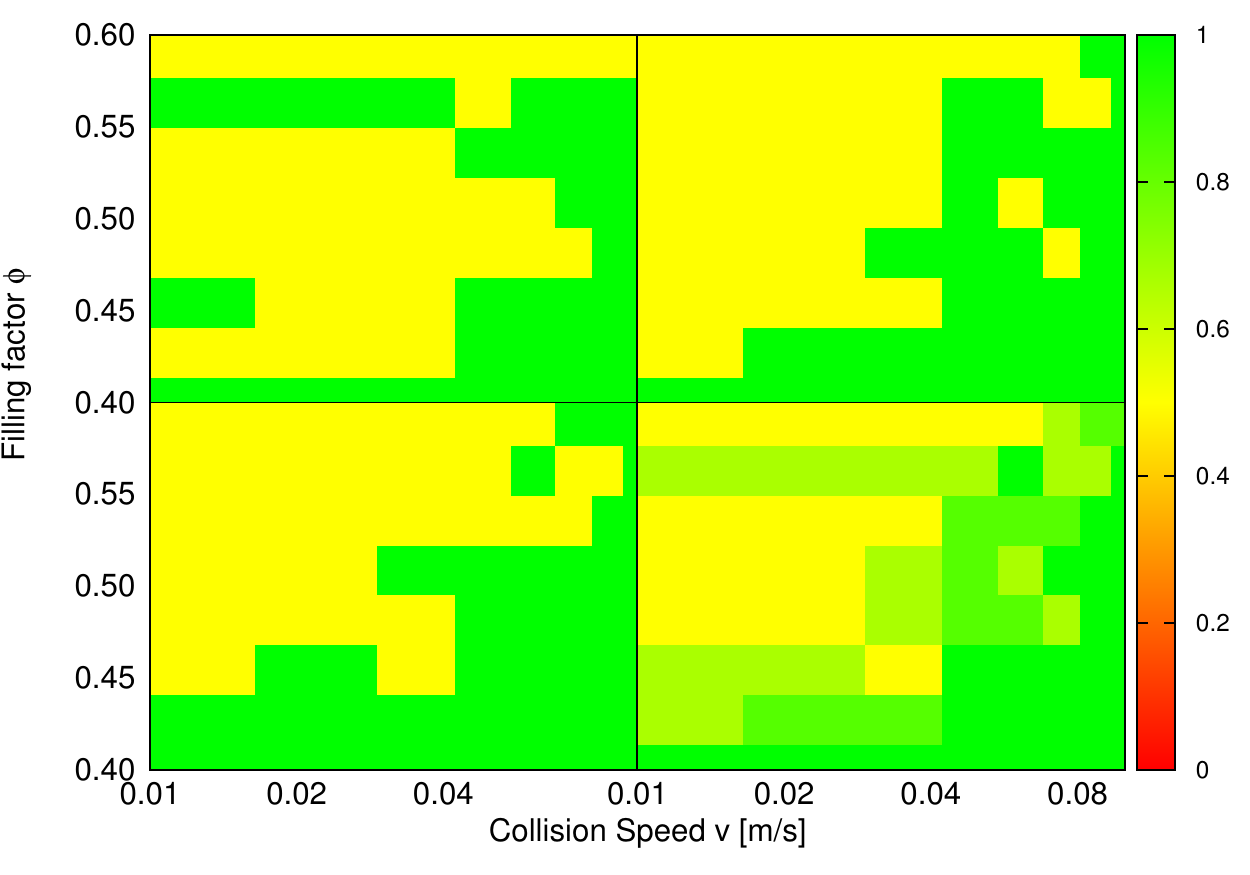}} \hfill \resizebox{0.50\hsize}{!}{\includegraphics{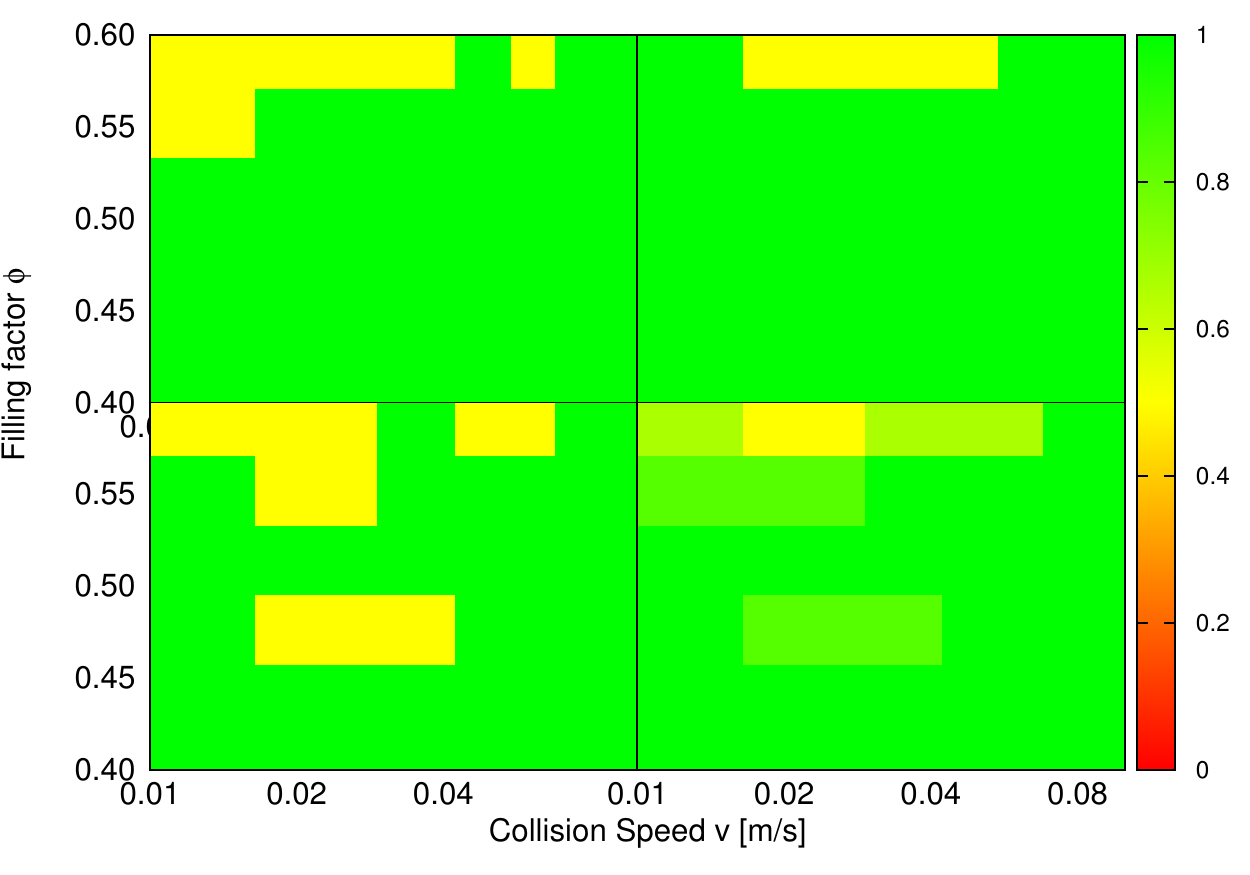}}
\caption{Growth factor of the collision of two aggregates with a diameter of $100\,\mathrm{\mu m}$ for three different orientations. The plot on the bottom right of both panels shows the values averaged over the three orientations. \textbf{Left:} CPE-aggregtes. \textbf{Right:} BAM aggregates (center migration).}
\label{fig:size_dependency}
\end{figure*}

To further examine the influence of the aggregate size we performed collisions of $100\,\mathrm{\mu m}$-sized CPE and BAM aggregates (using the center migration method as it yields the BAM aggregates with the highest filling factors). For the CPE aggregates we observe slightly more bouncing for filling factors between $0.4$ to $0.5$ (see left panel of Fig.\,\ref{fig:size_dependency}). However, for BAM aggregates there is no noticeable difference compared to the $60\,\mathrm{\mu m}$ aggregates (see right panel of Fig.\,\ref{fig:size_dependency}).

Depending on the filling factor the $100\,\mathrm{\mu m}$ aggregates consist of up to $350,000$ monomers. In order to analyze the size dependency it would be desirable to simulate collisions of even larger aggregates. Unfortunately, this is rendered impossible by the lack of available computing power. Simulating a single collision of two $100\,\mathrm{\mu m}$ sized aggregates took 10 to 20 hours (due to the different filling factors) on a GPU. Doubling the size would require computing times on the order of weeks for a single collision. For each orientation shown in Fig.\,\ref{fig:size_dependency} $48$ collisions have been performed.

\subsection{Impact parameter}
\label{sec:impact_parameter}

\begin{figure*}
\resizebox{0.50\hsize}{!}{\includegraphics{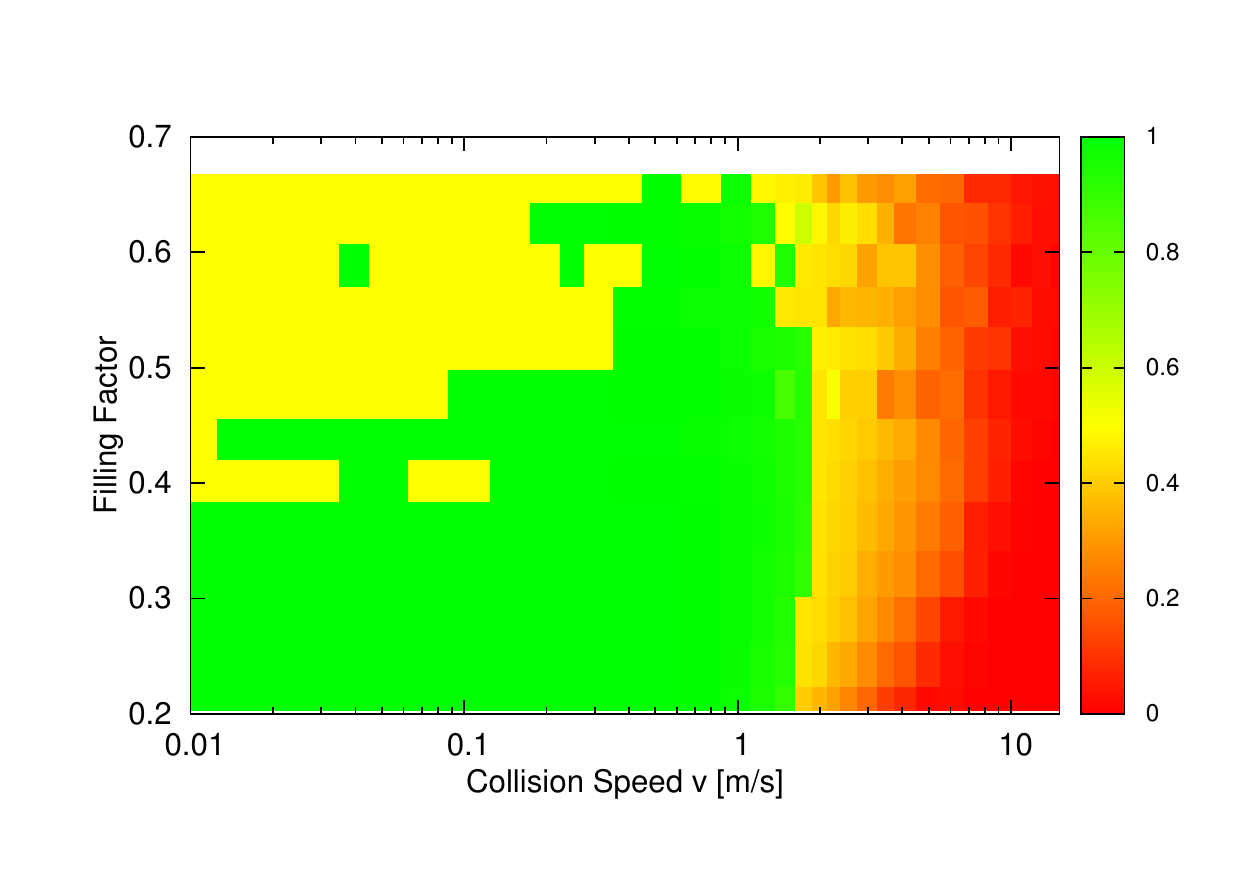}} \hfill \resizebox{0.50\hsize}{!}{\includegraphics{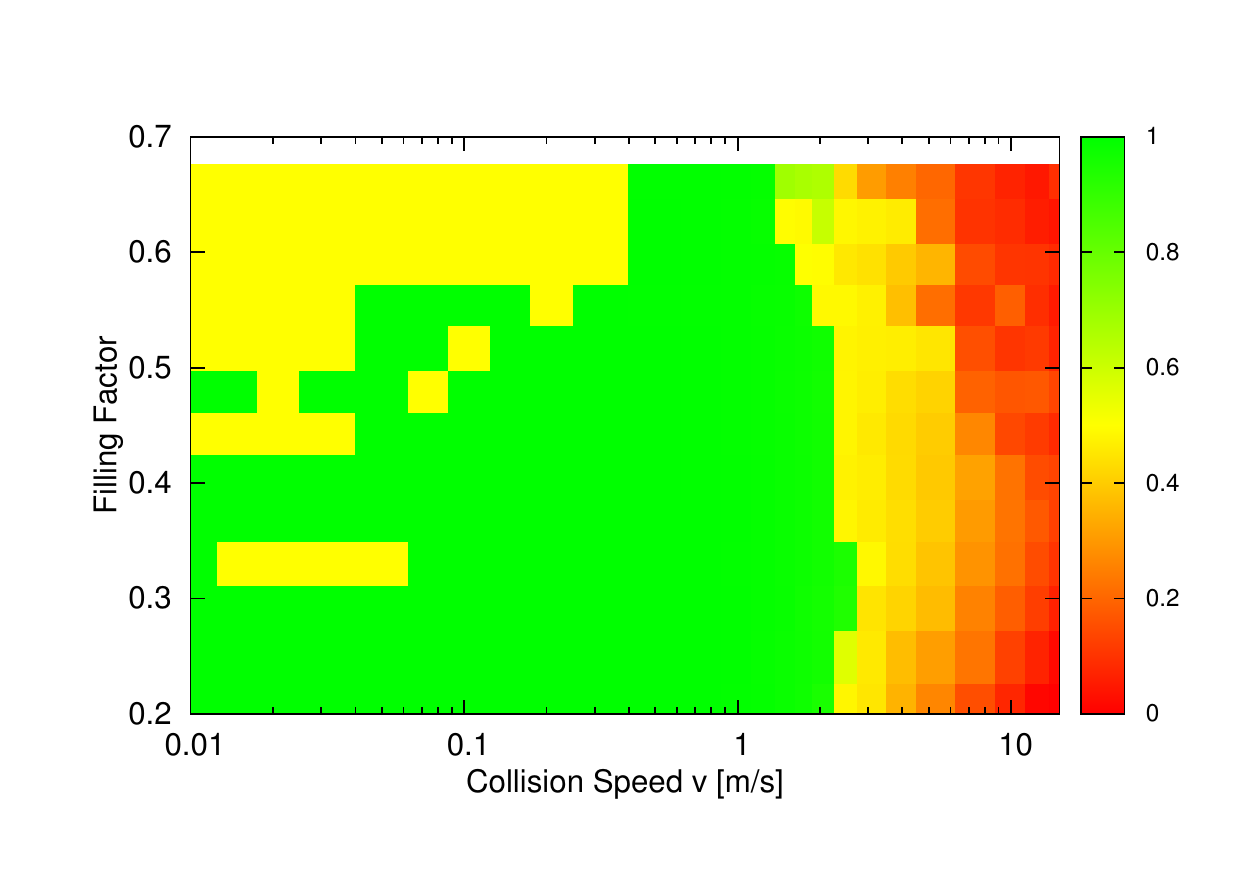}}
\caption{Growth factor of the collision of two CPE aggregates with an impact parameter $b = 0.5$. \textbf{Left:} Aggregates with a diameter of $30\,\mathrm{\mu m}$. \textbf{Right:}  Aggregates with a diameter of $60\,\mathrm{\mu m}$.}
\label{fig:impact_parameter_hex}
\end{figure*}

\begin{figure*}
\resizebox{0.50\hsize}{!}{\includegraphics{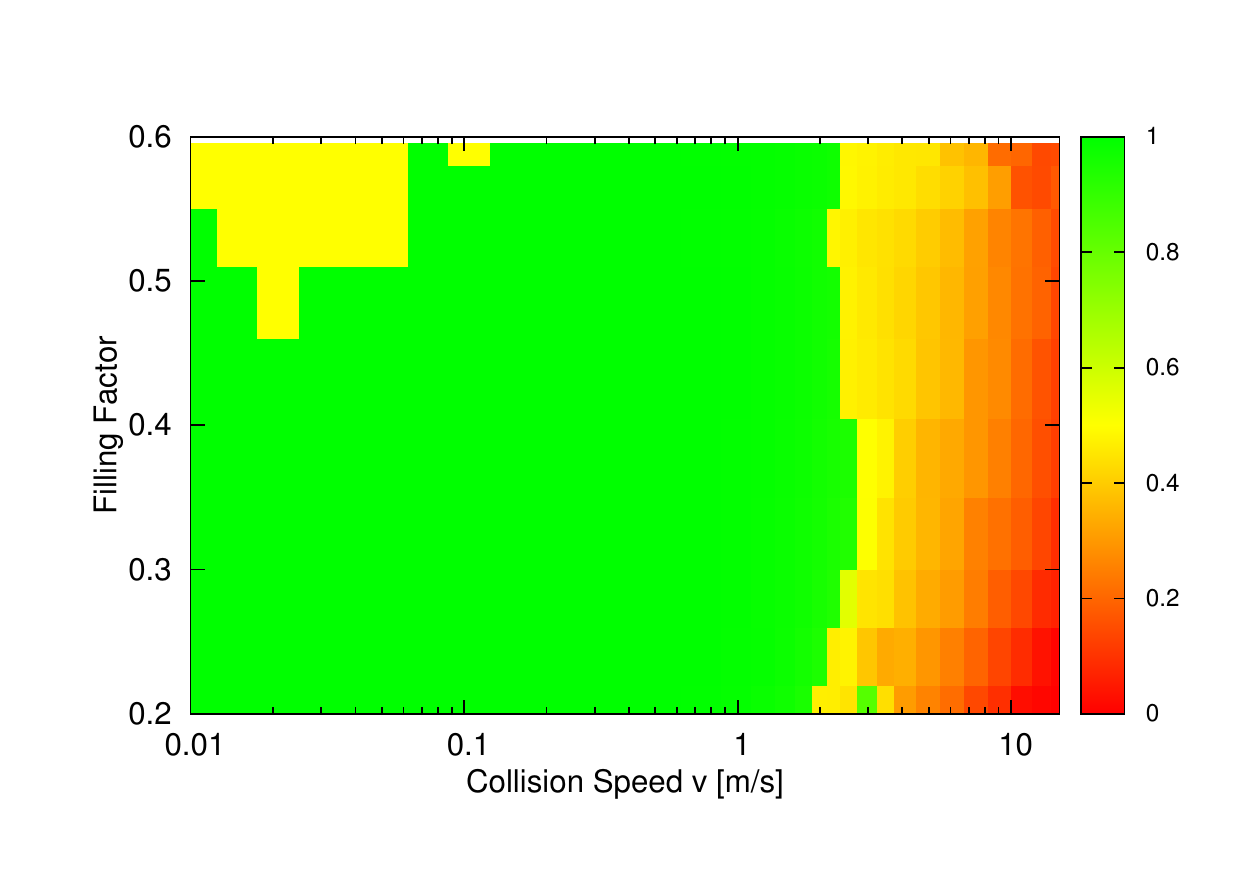}} \hfill \resizebox{0.50\hsize}{!}{\includegraphics{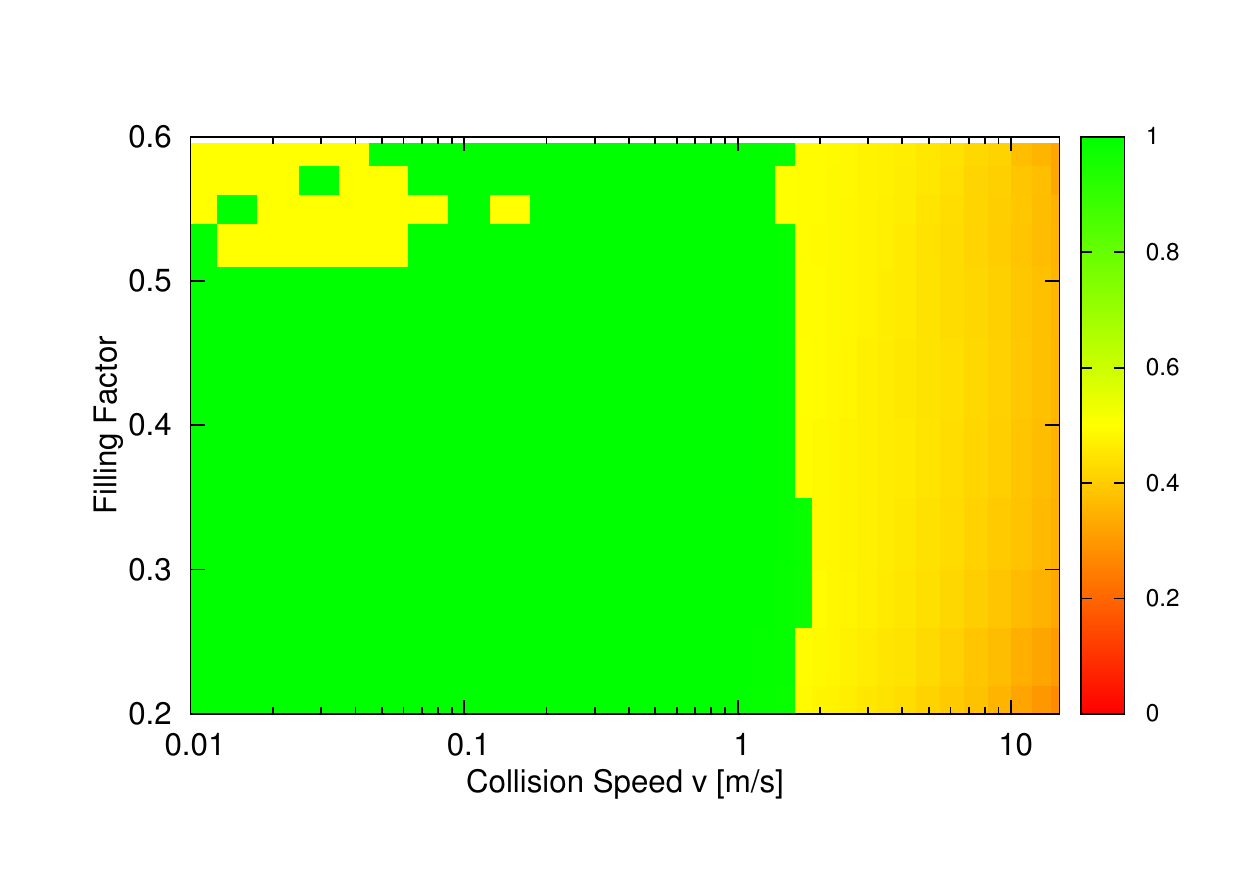}}
\caption{Growth factor of the offset collisions of two $60\,\mathrm{\mu m}$ sized BAM aggregates. \textbf{Left:} Impact parameter $b = 0.5$. \textbf{Right:} Impact parameter $b = 0.75$.}
\label{fig:impact_parameter_bam}
\end{figure*}

As a last step we examine the influence of the impact parameter $b = 0.5$. For the collisions, we used the same orientation as for the results shown in the right panel of Fig.\,\ref{fig:hex_orientation_dependency}. Contrary to our expectations we do not observe a significant influence of the impact parameter on the bouncing behavior of CPE aggregates (see Fig.\,\ref{fig:impact_parameter_hex}). However, fragmentation sets in at considerably lower velocities of $v_{\mathrm{s}\rightarrow\mathrm{f}} \approx 3\,\mathrm{m s^{-1}}$. In a head on collision the entire aggregate may dissipate the kinetic impact energy by internal restructuring and thus help to avoid fragmentation. This does not apply to offset collisions where it is easier to tear away the outer layers without major restructuring of the core of the aggregates.

However, comparing the growth map of the collisions between aggregates with a diameter of $30$ and $60\,\mathrm{\mu m}$ we do not observe any significant increase of velocity $v_{\mathrm{s}\rightarrow\mathrm{f}}$ where the transition from sticking to fragmentation occurs (see left and right panel of Fig.\,\ref{fig:impact_parameter_hex}). As already pointed out by \citet{2009ApJ...702.1490W} the increase of $v_{\mathrm{s}\rightarrow\mathrm{f}}$ for larger aggregates is limited to the case of head-on collisions.

In contrast, for BAM aggregates we do observe bouncing in a larger regime compared to case of head-on collisions (see Fig.\,\ref{fig:impact_parameter_bam}). Again, the same samples and orientation as for the head-on collisions (right panel of Fig.\,\ref{fig:bam_size_dependency}) have been used.

\section{Requirements for bouncing}

\begin{figure*}
\resizebox{0.50\hsize}{!}{\includegraphics{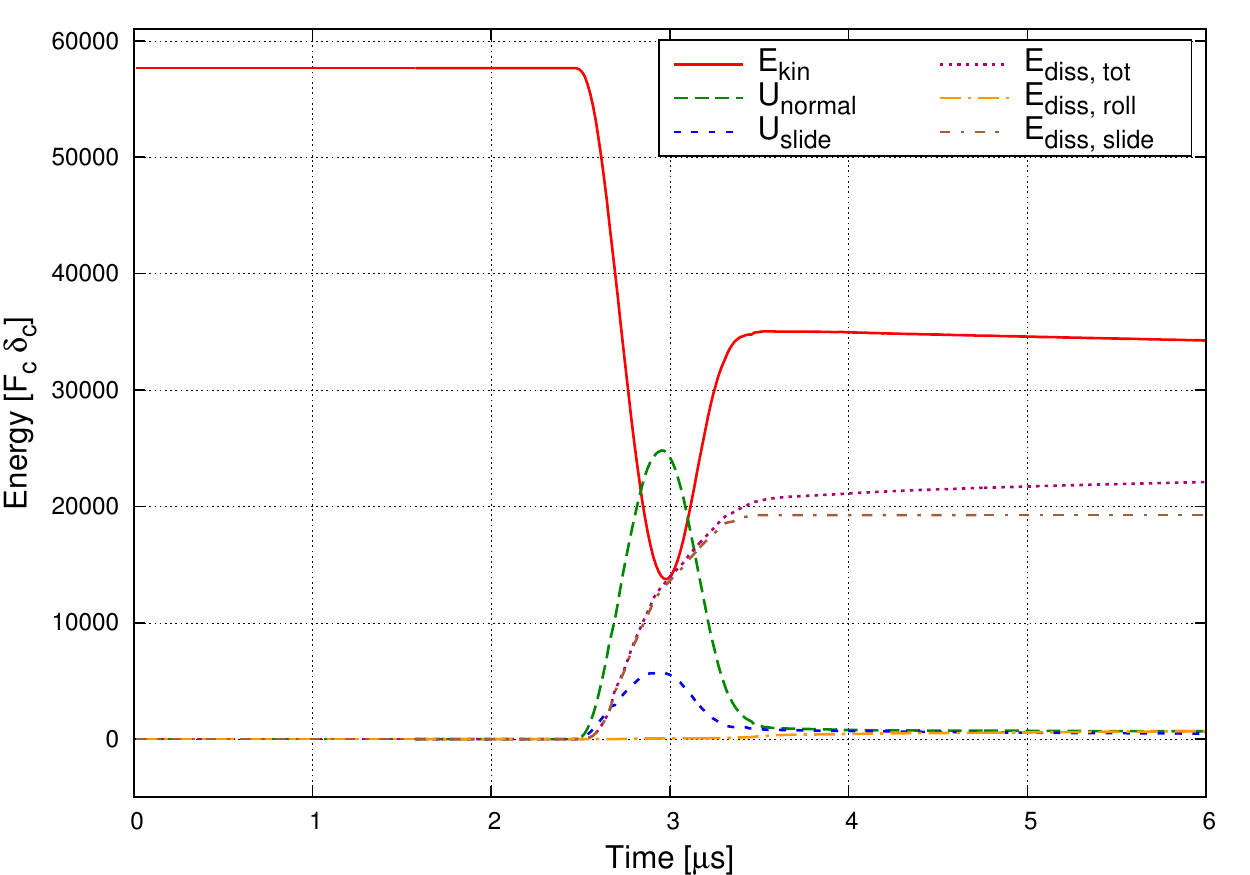}} \hfill \resizebox{0.50\hsize}{!}{\includegraphics{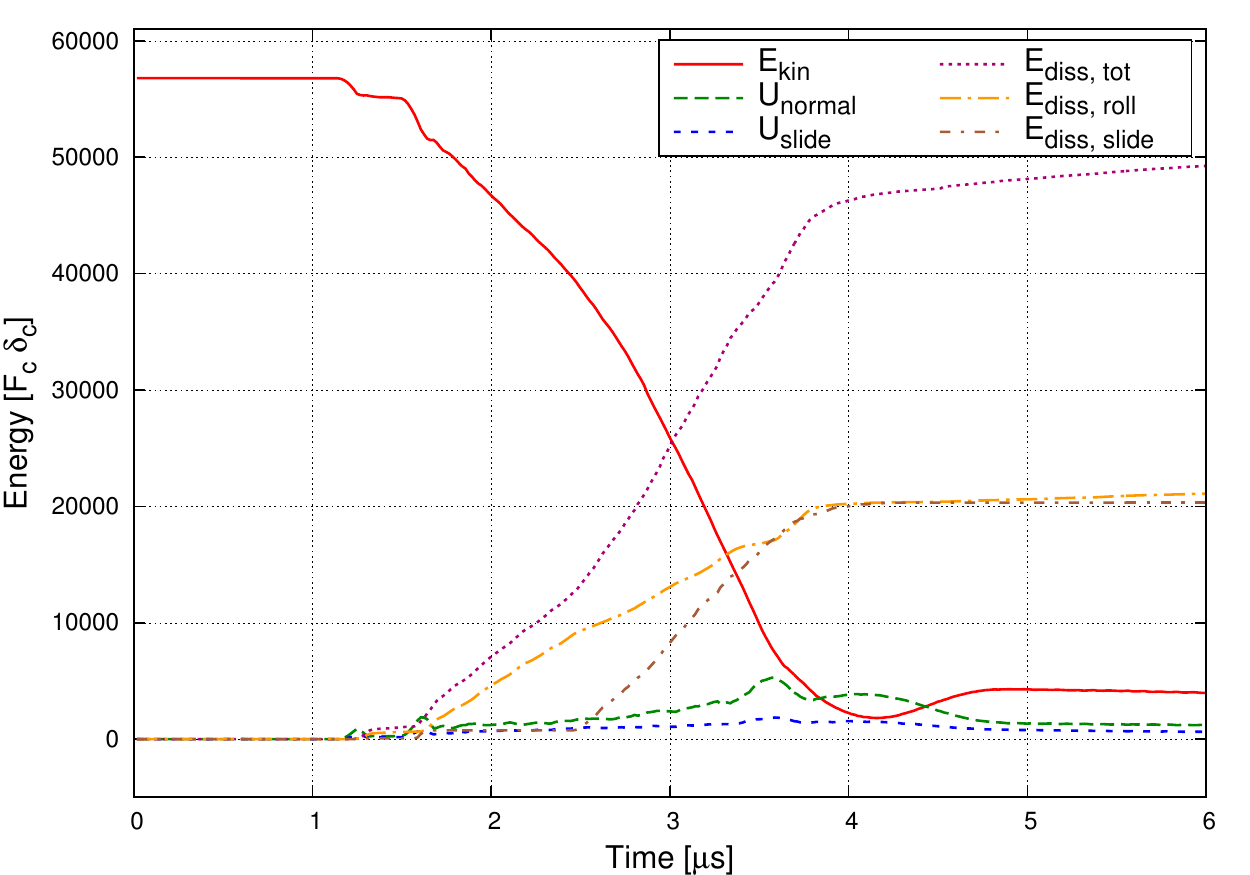}}
\caption{Time evolution of the kinetic, potential and dissipated energy during collisions of two aggregates with a diameter of $60\,\mathrm{\mu m}$ and filling factor $\phi \approx 0.59$. \textbf{Left:} A bouncing collision between two CPE aggregates. A sufficient amount of the impact energy is temporarily stored in the potential $U_{\mathrm{normal}}$ of the normal interaction. \textbf{Right:} During a sticking collision of two BAM aggregate most kinetic impact energy is dissipated.}
\label{fig:energies}
\end{figure*}

Compared to the aggregates generated by BAM or static compaction the bouncing regime of CPE aggregates is significantly larger. A likely explanation for this discrepancy is given by the different structure of the aggregates. To gain a deeper insight in the physical processes leading to sticking or bouncing it is worthwhile to have a closer look at a single collision.

Two aggregates may bounce off each other only if there is enough elastic energy left to break the contact area. Thus, a significant amount of the kinetic impact energy must be stored temporarily without being dissipated. If the colliding aggregates penetrate each other too deeply the impact energy is dissipated upon internal restructuring in the area where the contact is established. In this case inelastic sliding and rolling constitute the main dissipation channels \citep{2011ApJ...737...36W}. Thus, the ratio of elastic to dissipated energy of colliding aggregates is the key parameter that determines whether sticking or bouncing will occur.

Being able to track the evolution of the different types of energies over time is the key advantage of the model presented by \citet{2007ApJ...661..320W}. To address the different behavior of BAM and CPE aggregates we compare a bouncing collision of two CPE aggregates with a sticking collision of BAM aggregates. Both aggregates are $60\,\mathrm{\mu m}$ in diameter and have a filling factor $\phi \approx 0.59$. The time evolution of different types of energies and potentials for such collisions is shown in Fig.\,\ref{fig:energies}.

As expected, in the sticking case most kinetic energy is dissipated by inelastic sliding and rolling (right panel of Fig.\,\ref{fig:energies}). Only a small percentage of the impact energy is stored in the elastic regime of the normal $U_{\mathrm{normal}}$ and sliding potential $U_{\mathrm{slide}}$ (since the elastic energy stored in the rolling and twisting potentials is negligible they are not shown in Fig.\,\ref{fig:energies}).

We observe an entirely different situation in the bouncing case: As shown in the left panel of Fig.\,\ref{fig:energies} only about one third of the impact energy is dissipated, whilst roughly half of the kinetic energy is temporarily converted into potential energy. This coincides well with our predictions above. The potential energy that is stored mainly in the normal and sliding interaction is converted back into kinetic energy and allows that the colliding aggregates to separate again.

We can conclude that due to their lattice structure CPE aggregates can convert significantly more impact energy into elastic energy than BAM aggregates. In a compact CPE aggregate the monomers are located in densely packed layers. When the outer monomer of such a layer hits the other aggregate it is pushed inwards and will compress the layer. This way, kinetic energy is converted into potential energy without the occurrence of inelastic restructuring. This mechanism works well in the presence of a regular grid structure as it is the case for CPE aggregates. However, the monomers of BAM aggregates are not arranged in any regular pattern. Thus, they are not likely to bounce unless they are very compact in which case energy dissipation by internal restructuring is hindered because the monomers are locked in their position.

It also offers an explanation for the lower impact velocity of BAM aggregates at which we observe the transition from bouncing to sticking. At impact velocities above $0.1\,\mathrm{m s^{-1}}$ the kinetic impact energy is sufficient to restructure the monomers in the contact area that had been locked at lower impact velocities. Since the lattice structure of CPE aggregates offers higher resistivity against restructuring their transition velocity from bouncing to sticking is roughly $0.3\,\mathrm{m s^{-1}}$.

\section{Conclusions}
From the analysis of the statistical properties of the different samples presented in Sect.\,\ref{sec:sample_generation} we clearly see that the preparation method plays a crucial role when studying the collisional behavior of microscopic dust aggregates. As the relation between the volume filling factor and the average coordination number strongly depends on the preparation method one must be careful when comparing results obtained from computer simulations with laboratory work. In most laboratory experiments, only the volume filling factor can be measured, while the likewise important coordination number remains unknown. Since the more compact aggregates used in laboratory experiments are typically produced by mechanical compression, we expect that their microscopic structure resembles the static compaction aggregates much more closely than the hexagonal lattice type aggregates.

For computer simulations, generating large, compact aggregates by static compaction is infeasible due to the additional computational effort. We suggest to use BAM aggregates as an alternative. Their statistical properties are close to the aggregates produced by static compaction, yet they can be generated directly. Additionally, one does not run into problems caused by elastic charging as the generation procedure ensures that BAM aggregates are perfectly relaxed. At least in the various collisions simulations performed in this work BAM and static compaction aggregates show very similar behavior.

Based on the outcome of the simulations presented in this work we can conclude that bouncing collisions of dust aggregates in the size regime below $0.1\,\mathrm{mm}$ are rare. Unless the aggregates feature a regular lattice structure, which is not likely to be the case for the aggregates in a protoplanetary disk, bouncing requires filling factors greater than $0.5$ and collision velocities below $0.1\,\mathrm{m s^{-1}}$. Even if these prerequisites are met bouncing does not occur very frequently. Additionally, laboratory experiments on dust growth show that the maximum filling factor that is achieved during the growth process is much lower than $\phi = 0.5$ \citep[e.g.\,][]{2011ApJ...742....5T}. Therefore, the influence of bouncing on the growth process is limited in the sub-mm size regime.

Hence, from a microscopic view it remains unclear how cm-sized aggregates with filling factors considerably below $0.5$ are able to bounce off each other. The idea of a compacted outer layer (also referred to as a hard shell) has been put forward as a possible explanation. Using SPH-simulations, Geretshauser et al. (in preparation) observed that such a hard shell can indeed lead to bouncing collisions between aggregates with a porous core. \citet{2011ApJ...737...36W} obtained similar results when performing molecular dynamics simulations of collisions of CPE aggregates featuring a hard sphere. \citet{2008ApJ...675..764L} found that molding an aggregate significantly alters the outcome of a collision experiment. However, \citet{KotheInPrep} analyzed aggregates used in their collision experiments with X-ray computer tomography imaging and could not find any compacted outer layers.

In Sect.\,\ref{sec:impact_parameter} we have shown that offset collisions result in bouncing somewhat more often than head-on collisions. Depending on the experimental setup head-on collisions will be rare, and in a free collision with many particles setup head-on collisions will be rare as well \citep[e.g.\,][]{2012Icar..218..688W, 2012Icar..218..701B}. Thus, the impact parameter helps to resolve some of the discrepancies between numerical simulations and laboratory experiments.

Taking into account the different aggregate types we can only partially confirm the $n_\mathrm{c} \geq 6$ criterion for bouncing proposed by \citet{2011ApJ...737...36W}. It agrees well with our results from collisions of CPE aggregates. However, for BAM aggregates generated by the shortest migration method we observe very little bouncing at $n_\mathrm{c} = 6$ (see Sect.\,\ref{sec:results_bam} and Fig.\,\ref{fig:bam_shortest}). Likewise, for static compaction aggregates or BAM aggregates generated by the random or center migration method there are a few bouncing events where $n_\mathrm{c}$ is considerably lower than 6. A volume filling factor of $\phi \approx 0.5$ appears to constitute a lower limit for bouncing. At the present time we do not have an explanation what determines the exact value of the critical filling factor for the onset of bouncing. We have shown that it depends on the amount of energy that can be stored in the normal potential. The energy deposition is a continuous process, and it is to be expected that below a certain density sticking ensues. Numerically, we determined this value to be $\phi \approx 0.5$ in the sub-mm size regime.

Our simulations give insight into the fragmentation threshold as well. For small aggregates ($30\,\mathrm{\mu m}$) the fragmentation velocity is around $4\,\mathrm{m/s}$. Upon increasing the projectile size the fragmentation threshold rises up to about $10\,\mathrm{m/s}$ for the largest particle sizes we considered ($60\,\mathrm{\mu m}$), and this is independent of the sample generation method. This value is in very good agreement with the findings of SPH simulation for much larger objects \citep{2011A&A...531A.166G}. The shift to a larger fragmentation velocity is caused by the fact that larger particles can dissipate more energy than smaller particles. Upon increasing the filling factor, the fragmentation threshold decreases because the aggregates become much stiffer and cannot be deformed so easily. However, as shown in Sect.\,\ref{sec:impact_parameter} this effect only applies to the case of head-on collisions.

With respect to the growth of small dust agglomerates in the protoplanetary nebula our results indicate that for more realistic aggregates (BAM-type) bouncing only occurs for very small collision velocities ($< 0.1$ m/s) and large filling factors $>0.5$. Thus, the bouncing barrier may not be such a strong handicap in the growth phase of dust agglomerates, at least in the size range of $\approx 100\,\mathrm{\mu m}$. For larger, $m$-sized particles SPH results indicate bouncing up to $1\,\mathrm{m/s}$.

\begin{acknowledgements}
A.\,Seizinger acknowledges the support through the German Research Foundation (DFG) grant KL 650/16. The authors acknowledge support through DFG grant KL 650/7 within the collaborative research group FOR\ 759 {\it The formation of planets}.
We also thank the referee, Koji Wada, for his helpful comments that improved the quality of this paper.
A part of the simulations were performed on the bwGRiD cluster, which is funded by the Ministry for Education and Research of Germany and the Ministry for Science, Research and Arts of the state Baden-W\"urttemberg.
\end{acknowledgements}

\bibliographystyle{bibtex/aa}
\bibliography{bibtex/references}

\end{document}